\documentclass[sn-mathphys,Numbered]{sn-jnl}

\usepackage{caption}
\usepackage{tikz}
\usepackage{graphicx}%
\usepackage{multirow}%
\usepackage{physics}
\usepackage{amsmath,amssymb,amsfonts}%
\usepackage{amsthm}%
\usepackage{mathrsfs}%
\usepackage[title]{appendix}%
\usepackage{xcolor}%
\usepackage{textcomp}%
\usepackage{manyfoot}%
\usepackage{booktabs}%
\usepackage{algorithm}%
\usepackage{algorithmicx}%
\usepackage{algpseudocode}%
\usepackage{listings}%

\theoremstyle{thmstyleone}%
\theoremstyle{thmstyletwo}%

\theoremstyle{thmstylethree}%

\begin{document}

\title[The Other Side of the Coin]{The Other Side of the Coin: Recipient Norms and Their Impact on Indirect Reciprocity and
Cooperation}


\author*[1]{\fnm{Alina} \sur{Glaubitz}}\email{Alina.Glaubitz.GR@dartmouth.edu}

\author[1,2]{\fnm{Feng} \sur{Fu}}\email{Feng.Fu@dartmouth.edu}

\affil[1]{\orgdiv{Department of Mathematics}, \orgname{Dartmouth College}, \city{Hanover}, \postcode{03755}, \state{NH}, \country{USA}}

\affil[2]{\orgdiv{Department of Biomedical Data Science}, \orgname{Geisel School of Medicine at Dartmouth}, \city{Lebanon}, \postcode{03756}, \state{NH}, \country{USA}}


\abstract{Human cooperation depends on indirect reciprocity. In this work, we explore the concept of indirect reciprocity using a donation game in an infinitely large population. In particular, we examine how updating the reputations of recipients influences cooperation. Our work adds a  time-scale parameter for updating donor and recipient reputations. We find a trade-off between the level of cooperation and evolutionary stability influenced by social norms. `Forgiving' recipient norms enhance cooperation but increase susceptibility to defectors, whereas `unforgiving' norms reduce cooperation but defend against invasion by defectors. Expanding to include gossip groups allows us to analyze the evolutionary dynamics of the time-scale parameter, identifying `generous' norms that support cooperation, and `strict' norms
that discourage such generosity, ultimately showing vulnerability to defector invasions and potential cooperation collapse. }

\keywords{Reputation, Virtous Victim, Evolutionary Game Theory, Adaptive Dynamics}



\maketitle

\section{Introduction}

Cooperation is a fundamental phenomenon in the natural world. From the division of labor in ant colonies \cite{nowak2010eusocial} to the social hierarchies of apes \cite{Meder2013}, cooperation is all around us.
Due to the risk of exploitation and relative advantage of defectors, this has been a conundrum for researchers for a long time. \cite{nowak2006five} summarizes pivotal mechanisms that could promote the evolution of cooperation by facilitating assortment between cooperators: kin selection, direct reciprocity, \textbf{indirect reciprocity}, network reciprocity, group selection. Among these mechanisms, indirect reciprocity is arguably the most distinctly \textbf{human}, playing a crucial role in our societal evolution. Unlike chimpanzee groups, which number in the dozens, the scale and complexity of human cooperation is remarkable.

Recent studies in both human and animal societies have significantly advanced our understanding of cooperation and reciprocity. The prevalence of generalized reciprocity as well as direct and indirect reciprocity has been compared within human groups~\cite{STANCA2009190}. 
This study is complemented by explorations of the cognitive demands of various forms of reciprocity, underlining its widespread occurrence across species~\cite{Schweinfurth2019}. Insights into the developmental psychology of infants reveal a shift in their preference for 'always nice' individuals to those who rightfully punish 'bad' actors as they age~\cite{Hamlin2011}.  
While these studies focus on experimental settings, indirect reciprocity has also been evidenced in large-scale, real-world energy-saving behaviors~\cite{Yoeli2013}. Recent work on reputations has further found the 'virtuous victims' effect \cite{Jordan2020}, where crime victims gain higher reputations, a phenomenon exploited by individuals with Dark Triad personality traits~\cite{Ok2020SignalingVV}. 

These findings have been emphasized by advancements in mathematical modeling. See for example ~\cite{Okada2020, Nowak2005} for comprehensive reviews on the modeling of indirect reciprocity and its evolution. The concept of image scores laid the groundwork for understanding indirect reciprocity~\cite{Nowak1998}. Building upon this work, \cite{Leimar2001} shows the evolutionary instability of image scoring and discusses alternative strategies and social norms. This work has been furthered this by analyzing a binary reputation model, comparing all possible social norms and identifying eight evolutionarily stable social norms, known as the \textbf{leading eight}~\cite{Ohtsuki2004}.

The leading eight model relies on shared reputations within the population, but there have been advances to asess private reputations~\cite{Uchida2010,Ohtsuki2015,Hilbe2018} that have shown that the assumption of shared reputations is crucial for the Leading Eight model. Recent work further extends this framework to consider generous reputation updates~\cite{Schmid2021TheEO} and the impact of the desire to please others~\cite{Krellner2021}. Additional work has included private reputation updates along with higher-order moral assessment \cite{Perret2021}, the reputation structure under private assessment \cite{Fujimoto2022} as well as shown the evolutionary stability of cooperation under private assessment \cite{fujimoto2023evolutionary,Schmid2023}.

The original framework has been extended by relaxing the assumption of binary reputation updates. It has been shown that an extension to continuous rather than discrete social norms, can extend the influence of factors like ecological optima and individual preferences rather than historical precedence~\cite{Yan2023}. Additionally, this understanding has been extended by exploring local stability and ternary reputation updates of discrete reputations~\cite{Lee2021,Murase2022}.

Beyond these developments, our understanding of indirect reciprocity has been expanded through investigations into information sharing \cite{Nakamura2012}, empathy \cite{Radzvilavicius2019}, the complexity of social norms \cite{Santos2021}, and dual reputation updates~\cite{Murase2023}. Another approach has been adopted by comparing the Leading Eight strategies to each other in mixed population to compare the evolutionary stability against each other~\cite{Kessinger2023} as well as in conjunction with other forms of reciprocity ~\cite{Schmid2021TheEO,sasaki2023evolution}.

Our work builds upon these foundations, in particular dual reputation updates, introducing a third time-scale parameter to examine the balance between recipient and donor reputation updates as introduced in~\cite{Murase2023}. (\cite{Murase2023} considered an equal likelihood of dual update, yet the relative time scales of dual updates have remained answered.) We use a framework similar to the analysis in \cite{Kessinger2023} to compare different time-scales in this context. Our findings indicate that populations tend to update recipient norms to some extent, with different dynamics leading either to a cooperative equilibrium or a bistable state where cooperation may collapse. This challenges the notion from \cite{Ohtsuki2004} that judgments of good people cooperating with other good people are inconsequential. Our results suggest that how these individuals are judged is crucial and can have profound implications for the stability of cooperative systems.

\section{Model}

players are paired at random to play a game of the prisoner's dilemma in a well-mixed infinitely large population. 
Each player can cooperate with their partner (they can pay cost $c>0$ for their partner to receive benefit $b>c>0$) or defect. 
Each player either has a good or a bad reputation that is updated based on their interactions with other players. 
We are interested in understanding if the reputation updates should only concern the player's own actions or also the actions of their co-players (as observed by \cite{Jordan2020}). 
To address this question we use the framework introduced by \cite{Ohtsuki2004}. 
We assume that there is a separation of time-scales: on a shorter time-scale, reputations are updated. 
Once they are in equilibrium, players update their strategies on a longer time-scale. 
We introduce a third time-scale $q$ that regulates the frequency of donor updates compared to recipient updates. 
For this model, the social norms are fixed. 
Further, we assume an indirect observation model. 
For each interaction, a third player observes the interaction and shares her updated opinions with the rest of the population. 
This way, everyone in the population shares the same opinion even in the presence of errors. 
Under these assumptions, we can analyze the model to find evolutionarily stable pairs of social norms and behavioral strategies (ESSes) for a fixed ratio of updating based on donation behavior $q$. 
Once, we finished this analysis, we focus on the emerging ESSes to gain a better understanding of the time-scale value $q$. 
Particularly, we introduce a third time-scale. 
Once the population is in an ESS, players can adapt the time-scale value $q$. 

\begin{figure}[tb]
    \centering
    \includegraphics[width=\textwidth]{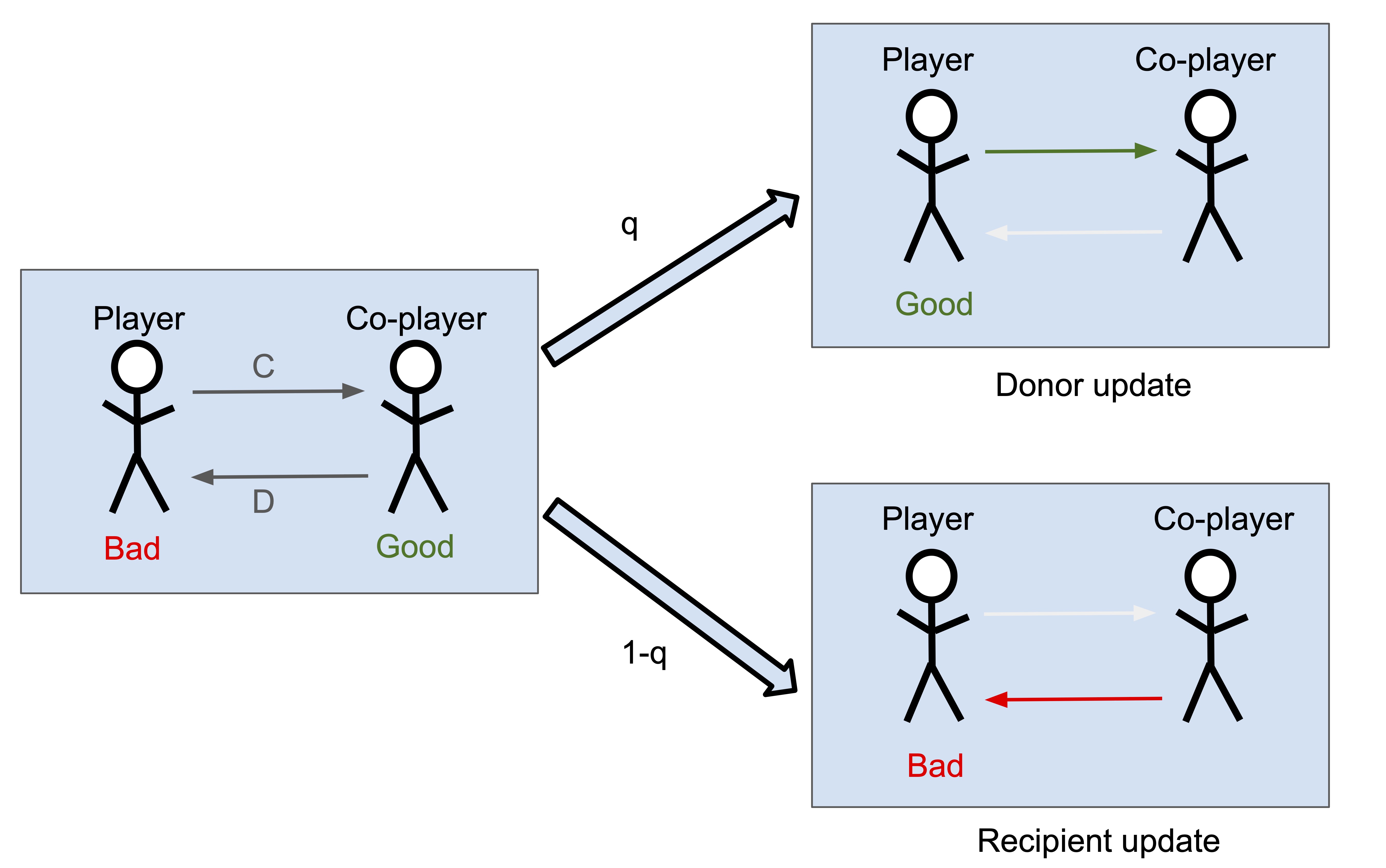}
    \caption{Illustration of possible reputation updates. In each interaction, with rate $q$, the reputation of the donor is updated and with rate $(1-q)$ the reputation of the recipent is updated.}
    \label{fig:ESS}
\end{figure}

\newpage

\subsection{Reputation Dynamics}

To find evolutionarily stable strategies, we fix social norms as described in Table \ref{tb:norms} and and assume that all players follow the same behavioral strategy of the form described in Table \ref{tb:beh}. Each social norm describes how to update a player's reputation in each possible scenario. E.\,g.\ $d_{\text{GBD}}^{(R)} = G$ means that if I have a good reputation and interact with a bad player who defects against me I obtain a good reputation. Similarly, the behavioral strategy determines the player's behavior in each scenario. E.\,g.\ $p_{\text{GB}}=D$ means that a good player will defect against a bad player.

\begin{table}[tb]
 \begin{tabular}{||c|c||c c c c||} 
 \hline
 \textbf{Role}& \textbf{Action} & \textbf{BB} & \textbf{BG} & \textbf{GB} & \textbf{GG} \\ [0.5ex] 
 \hline\hline
\textbf{Donor} & \textbf{Defect} & $d_{\text{BBD}}^{(D)}$ & $d_{\text{BGD}}^{(D)}$  & $d_{\text{GBD}}^{(D)}$  & $d_{\text{GGD}}^{(D)}$  \\
 \textbf{Donor} & \textbf{Cooperate} & $d_{\text{BBC}}^{(D)}$ & $d_{\text{BGC}}^{(D)}$  & $d_{\text{GBC}}^{(D)}$  & $d_{\text{GGC}}^{(D)}$ \\
 \textbf{Recipient} & \textbf{Defected against} & $d_{\text{BBD}}^{(R)}$ & $d_{\text{BGD}}^{(R)}$  & $d_{\text{GBD}}^{(R)}$  & $d_{\text{GGD}}^{(R)}$  \\
 \textbf{Recipient} & \textbf{Cooperated with} & $d_{\text{BBC}}^{(R)}$ & $d_{\text{BGC}}^{(R)}$  & $d_{\text{GBC}}^{(R)}$  & $d_{\text{GGC}}^{(R)}$ \\ [0.5ex] 
 \hline
 \end{tabular}
 \caption{Social Norms. For each possible interaction and updating role, we assign a new reputation $d_{\text{IJK}}^{(L)} \in \{G,B\}$. } \label{tb:norms}
 \end{table}

\begin{table}[tb]
 \centering
 \begin{tabular}{||c||c c||} 
 \hline
  & \textbf{B} & \textbf{G} \\ [0.5ex] 
 \hline\hline
 \textbf{B} &  $p_{\text{BB}}$ & $p_{\text{BG}}$  \\
 \textbf{G} & $p_{\text{GB}}$ & $p_{\text{GG}}$ \\
 \hline
 \end{tabular}

 \begin{tikzpicture}[remember picture,overlay]
   \node[anchor=north west,xshift=5cm, yshift=-3.75cm] at (current page.north west) 
   {\parbox{0.8\textwidth}{
    \captionof{table}[Behavioral Strategies]{Behavioral Strategies. Each player has a fixed behavioral strategy $p_{\text{IJ}} \in \{C,D\}$ when interacting with other players, depending on each player's reputation. \textcolor{white}{\\}}
    \label{tb:beh}
   }};
 \end{tikzpicture}
\end{table}

Then, we compare the payoffs of mutant strategies to understand which pairs of social norm and behavioral strategy $(d,p)$ are evolutionarily stable.

\textcolor{white}{hello \\ this is war \\}
We use the same reputation dynamics framework as \cite{Ohtsuki2004}, and in particular, we investigate the dynamics of the reputation scores. 
When $h_t(p',p)$ denotes the proportion of good $p'$ players in a population of $p$ players at time $t$, the dynamics of the reputation score in time are given by the ordinary differential equation (ODE)
\begin{align*}
        \frac{d}{dt}h_t(p',p) =h_t(p')&[h_t(p)(q d_{\text{GG}p_{\text{GG}}'}^{(D)}+(1-q)d_{\text{GG}p_{\text{GG}}}^{(R)})+(1-h_t(p))(q d_{\text{GB}p_{\text{GB}}'}^{(D)}+(1-q)d_{\text{GB}p_{\text{BG}}}^{(R)})] \\
        + (1-h_t(p'))&[h_t(p)(q d_{\text{BG}p_{\text{BG}}'}^{(D)}+(1-q) d_{\text{BG}p_{\text{GB}}}^{(R)})+(1-h_t(p))(q d_{\text{BB}p_{\text{BB}}'}^{(D)}+(1-q) d_{\text{BB}p_{\text{BB}}}^{(R)})] -h_t(p')
\end{align*}
By introducing small errors in assignment (with rate $\mu_a$) and execution (with rate $\mu_e$), we get
$$ D_{ij}(p',p) = ((1-2\mu_a)\{(1-\mu_e) (q d_{ijp_{ij}'}^{(D)}+(1-q) d_{ijp_{ji}}^{(R)})+\mu_e(q d_{ijD}^{(D)}+(1-q) d_{ijD}^{(R)})\} + \mu_a),$$
and therefore
\begin{align} \label{eq:dyn}
        \frac{d}{dt}h_t(p',p) &= h_t(p')[h_t(p)D_{\text{GG}}(p',p)+(1-h_t(p))D_{\text{GB}}(p',p)] \\
        &+ (1-h_t(p'))[h_t(p)D_{\text{BG}}(p',p)+(1-h_t(p))D_{\text{BB}}(p',p)] -h_t(p'). \nonumber
\end{align}
Similarly to \cite{Ohtsuki2004}, this ODE has a unique equilibrium $h_*(p',p)$. So, independent of the initial condition, the solution converges to this unique equilibrium. 
After long enough time, the probability of a resident player $p$ to receive a donation $b$ is given by
$$ \theta(p,p) = h^2_*(p,p) p_{\text{GG}} + h_*(p,p)(1-h_*(p,p))(p_{\text{GB}}+p_{\text{BG}}) + (1-h_*(p,p))^2 p_{\text{BB}}.$$
Similarly, the probability of a mutant player $p'$ to receive a donation $b$ is
\begin{align*}
    \theta(p,p') &= h_*(p',p)h_*(p,p) p_{\text{GG}} + h_*(p,p)(1-h_*(p',p))p_{\text{GB}}\\&+h_*(p',p)(1-h_*(p,p))p_{\text{BG}} + (1-h_*(p',p))(1-h_*(p,p)) p_{\text{BB}},
\end{align*}
and the probability of a mutant player $p'$ to donate at cost $c$ is 
\begin{align*}
    \theta(p',p) &= h_*(p',p)h_*(p,p) p_{\text{GG}}' + h_*(p,p)(1-h_*(p',p))p_{\text{BG}}'\\&+h_*(p',p)(1-h_*(p,p))p_{\text{GB}}' + (1-h_*(p',p))(1-h_*(p,p)) p_{\text{BB}}'.
\end{align*}
Therefore, the expected payoff of a resident $p$ is given by
$$ W(p|p) = \theta(p,p)(B-C), $$ while the expected payoff of a mutant $p'$ is
$$ W(p'|p) = \theta(p,p')B - \theta(p',p)C.$$
By comparing each of the 16 behavioral strategies $p$ to each possible mutant $p'$, we obtain evolutionarily stable pairs of social norms and strategies $(d,p)$.

To compare the level of cooperation, we compare the expected payoffs $W(p,p)$ to the maximal payoff that's possible $(B-C)$ in a fully cooperative population. The level of cooperation is given by $L_C=\theta(p,p)$.

For $\frac{b}{c}>1/q+\mathcal{O}(\mu)$, we find that $2^{3+6}=512$ social norms (in combination with one of the behavioral strategies) out of $2^{16}$ social norms form ESS pairs and achieve levels of cooperation that satisfy $L_C \geq 1 - \mathcal{O}(\mu)$, i.\,e.\ $L_C$ approaches 1 (or full cooperation) as the term $\mu$ becomes negligibly small.
For each of these social norms, the reputation updates for donors are the same as for one of the leading eight. So, we obtain $3$ free variables for that. 
Moreover, two more variables within the recipient norms are fixed: $d_{\text{GGC}}^{(R)} = G$ and $d_{\text{GBC}}^{(R)}=G$. 
In particular, the resulting highly cooperative ESS (CESS) social norms are of the form given in Table \ref{tb:hc}. 
Additionally, $d_{\text{BGD}}^{(R)}$ introduces a trade-off. 
Here, $d_{\text{BGD}}^{(R)}=G$
obtains higher levels of cooperation by forgiving execution errors after being punished by a cooperator, while also making higher benefit-to-cost ratios necessary, as cheaters (AllD) as cheaters are forgiven for their misdeeds without apology. 
On the other hand, $d_{\text{BGD}}^{(R)}=B$ obtains lower levels of cooperation, but is more stable against invasion by defectors.

\begin{table}[h!]
 \begin{tabular}{||c|c||c c c c||} 
 \hline
 \textbf{Role}& \textbf{Action} & \textbf{BB} & \textbf{BG} & \textbf{GB} & \textbf{GG} \\ [0.5ex] 
 \hline\hline
\textbf{Donor} & \textbf{Defect} & *  & B  & G  & B  \\
 \textbf{Donor} & \textbf{Cooperate} & *  & G  & *  & G  \\
 \textbf{Recipient} & \textbf{Defected against} & *  & \textcolor{red}{*}  & *  & *  \\
 \textbf{Recipient} & \textbf{Cooperated with} & *  & *  & G  & G  \\ [0.5ex] 
 \hline
 \end{tabular}

 \begin{tikzpicture}[remember picture,overlay]
   \node[anchor=north west,xshift=5.5cm, yshift=-14.65cm] at (current page.north west) 
   {\parbox{0.8\textwidth}{
    \captionof{table}[]{Cooperative social norms. Black values $\in \{G,B\}$ are fixed for all highly cooperative social norms. Stars are open. The red star \textcolor{red}{*} denotes the presence of a trade-off. For \textcolor{red}{*}=G, cooperation levels are higher, while for \textcolor{red}{*}=B, smaller benefit-to-cost ratio is necessary to achieve cooperation. \textcolor{white}{\\}}
    \label{tb:hc}
   }};
 \end{tikzpicture}
 \end{table}

\textcolor{white}{. \\ . \\ . \\}
 
To prove that any social-norm/behavioral-strategy pair $(d,p)$ satisfying 
\begin{itemize}
    \item[(1)] $p$ is evolutionarily stable under $d$
    \item[(2)] $L_C \geq 1 - \mathcal{O}(\mu)$
\end{itemize}
satisfies this property, we use a similary strategy to \cite{Ohtsuki2004}. 
In order to satisfy both (1) and (2), $(d,p)$ needs to maintain cooperation in the absence of errors. 
In this case, the reputation dynamics are given by 
\begin{align*}
    \frac{d h}{d t} &= \frac{1}{2}(h^2 ( d_{\text{GG}p_{\text{GG}}}^{(D)}+ d_{\text{GG}p_{\text{GG}}}^{(R)}) + h (1-h) ( d_{\text{GB}p_{\text{GB}}}^{(D)}+ d_{\text{GB}p_{\text{BG}}}^{(R)}+ d_{\text{BG}p_{\text{BG}}}^{(D)}+ d_{\text{BG}p_{\text{GB}}}^{(R)})\\& + (1-h)^2 ( d_{\text{BB}p_{\text{BB}}}^{(D)}+ d_{\text{BB}p_{\text{BB}}}^{(R)})) - h=: f(h).
\end{align*}
Then, $h=1$ needs to be a stable equilibrium. 
Hence, $f(1) = 0$ and $f'(1) \leq 0$ need to be satisfied. 
In particular, $f(1) = \frac{1}{2}( d_{\text{GG}p_{\text{GG}}}^{(D)}+ d_{\text{GG}p_{\text{GG}}}^{(R)}) - 1$ and therefore $ d_{\text{GG}p_{\text{GG}}}^{(D)} =  d_{\text{GG}p_{\text{GG}}}^{(R)} = 1$ needs to be satisfied. 
By applying second order stability analysis, we can also get $( d_{\text{BB}p_{\text{BB}}}^{(D)}+ d_{\text{BB}p_{\text{BB}}}^{(R)}) \geq 2$ (adjust for the correct formulation). 
There are a few possibilities here, to obtain stability. 
These cases are given by
\begin{itemize}
    \item[(1)] $( d_{\text{GB}p_{\text{GB}}}^{(D)}+ d_{\text{GB}p_{\text{BG}}}^{(R)}+ d_{\text{BG}p_{\text{BG}}}^{(D)}+ d_{\text{BG}p_{\text{GB}}}^{(R)}) = 4$
    \item[(2)] $( d_{\text{GB}p_{\text{GB}}}^{(D)}+ d_{\text{GB}p_{\text{BG}}}^{(R)}+ d_{\text{BG}p_{\text{BG}}}^{(D)}+ d_{\text{BG}p_{\text{GB}}}^{(R)}) = 3$
    \item[(3)] $( d_{\text{GB}p_{\text{GB}}}^{(D)}+ d_{\text{GB}p_{\text{BG}}}^{(R)}+ d_{\text{BG}p_{\text{BG}}}^{(D)}+ d_{\text{BG}p_{\text{GB}}}^{(R)}) = 2$
\end{itemize}
Now, in case (3), stability can only be obtained by second order terms, as $f'(1) = 0$. So, we need $d_{\text{BB}p_{\text{BB}}}^{(D)}+ d_{\text{BB}p_{\text{BB}}}^{(R)} \geq 1$ for $h=1$ to be a stable equilibrium. However, because of this second-order stability, small errors of size $\mu$ cause deviations from full cooperation at the order of $\mathcal{O}(\sqrt{\mu})$. On the other hand, for case (1) and case (2), stability is achieved by first-order terms. In this case, errors of size $\mu$ only cause a deviation at the order of $\mathcal{O}(\mu)$. What distinguishes these two cases is the eigenvalue of the linearized system. In case (2), the eigenvalue is given by $f'(1) = -\frac{1}{2}$, while in case (3), it is given as $f'(1)=-1$. So, in the presence of small errors, we achieve the highest levels of cooperation in case (1). In particular, we need to achieve the highest values of cooperation, we need $d_{\text{GB}p_{\text{GB}}}^{(D)} = 1$, $d_{\text{GB}p_{\text{BG}}}^{(R)} = 1$, $d_{\text{BG}p_{\text{BG}}}^{(D)} = 1$ and $d_{\text{BG}p_{\text{GB}}}^{(R)} = 1$. So, we only need to figure out what $p_{\text{GB}}$ and $p_{\text{BG}}$ should look like.

In what follows, $p_{\text{GB}} = D$ has to hold, in order to ensure that players with a bad reputation will be punished. Otherwise, players that always defect can invade the population, as they are gain the benefit from cooperation, while not paying the cost of cooperating. Therefore, $d_{\text{GBD}}^{(D)} = 1$ and $d_{\text{BGD}}^{(R)} = 1$ have to hold. Furthermore, it is necessary that $p_{\text{BG}} = C$. Thus, $d_{\text{BGC}}^{(D)} = 1$ and $d_{\text{GBC}}^{(R)} = 1$ have to hold. Finally, all of the previous discussion does not take into account that errors occur most frequently when two good players interact. So, in order to ensure the highest level of cooperation, we need to fix $d_{\text{GGD}}^{(R)} = 1$.

What might be most surprising about these findings is the fact that $d_{\text{BGD}}^{(R)} = 1$ increases the level of cooperation. 
This means that, if I have a bad reputation and someone defects against me, I will gain a good reputation, without ever cooperating/apologizing. 
Now, this is also the reason, why we need $\frac{b}{c} > 1/q + \mathcal{O}(\mu/q)$ in order for these social norms to be evolutionary stable. 
For $b\leq c/q$, AllD can invade a population using social norm $d$ for any other behavioral strategy, as AllD players obtain after time $1/q$ without ever cooperating. 
See Figure \ref{fig:ESS} for an illustration of this. 
In particular, we see here that even without errors, with probability $q$, AllD player have a good reputation (from being defected against). 
So, with probability $q$ conditional cooperators --- with strategy $p=\begin{pmatrix}
    0 & 1 \\ 0 & 1
\end{pmatrix}$ (OR) for social norms where $d_{\text{BBC}}^{\text{(D)}} = G$ and $d_{\text{BBD}}^{\text{(B)}} = G$ respectively $p=\begin{pmatrix}
    1 & 1 \\ 0 & 1
\end{pmatrix}$ (CO) players for the remaining social norms --- cooperate with AllD players. 
In particular, the payoff of AllD players becomes $qb$, while the payoff of CO/OR players is $(b-c)$. 
So, in order to avoid invasion by AllD players, $qb<(b-c)$ needs to be satisfied, which is equivalent to $b>c/q$. We obtain the bound $\frac{b}{c} > 1/q + \mathcal{O}(\mu/q)$ from including errors.

\begin{figure}[tb]
    \centering
    \includegraphics[width=\textwidth]{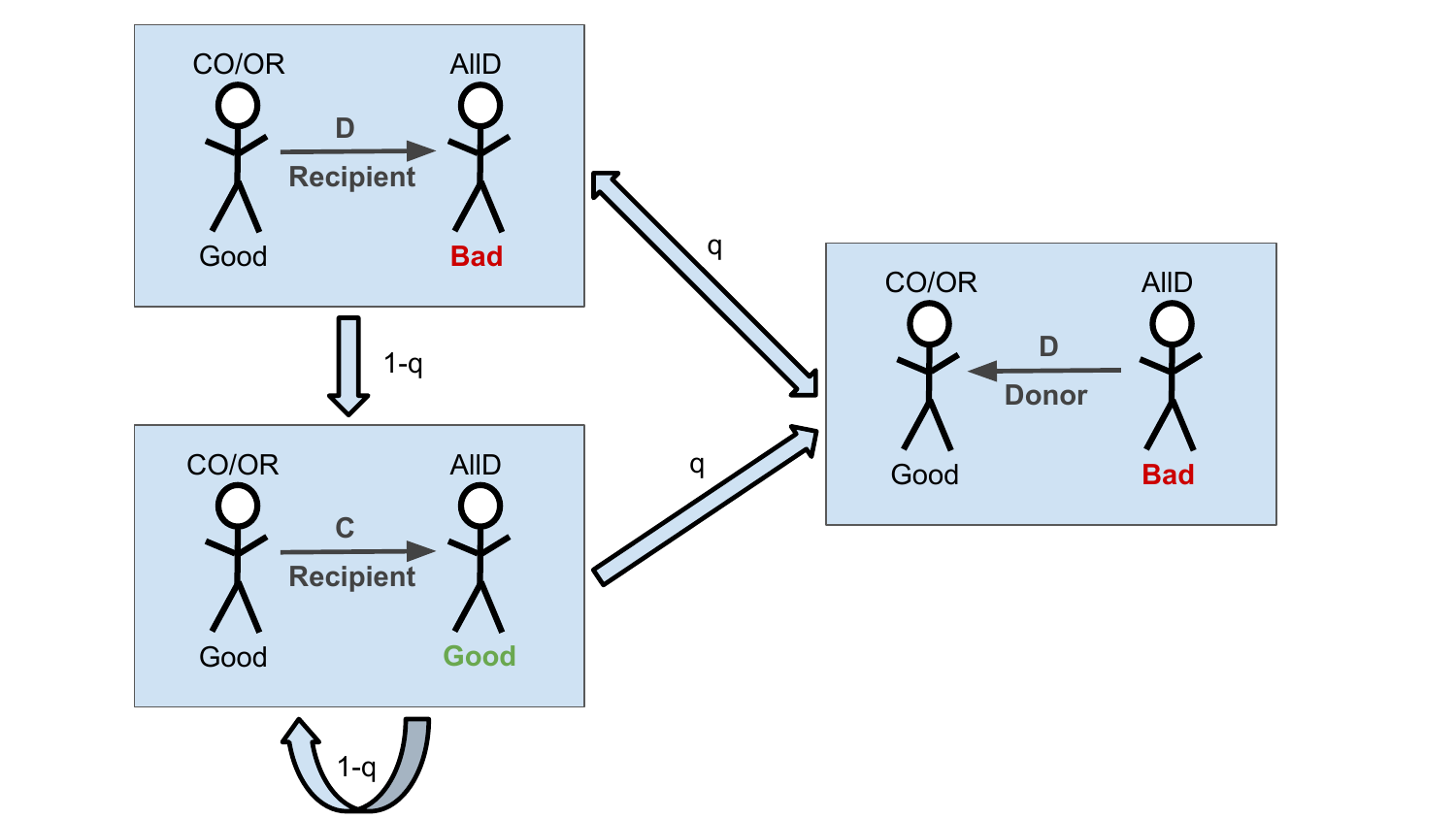}
    \caption{Conditional Cooperation (CO/OR) can only be an ESS if the benefit/cost ratio is greater than $\frac{1}{q} + \mathcal{O}(\mu).$ In this illustration we see that, when updating the reputation of donors with rate $q$ and the reputation of recipients with rate $(1-q)$, AllD players have a good reputation with rate $(1-q)$ and a bad reputation with rate $q$.}
    \label{fig:ESS}
\end{figure}

To summarize, besides the five variables that are fixed for donor reputation updates, there are three fixed variables in the Recipient social norms that achieve the highest level of cooperation in this model: 
\begin{itemize}
    \item[(1)] $d_{\text{BGD}}^{(R)} = G$ 
    \item[(2)] $d_{\text{GGC}}^{(R)} = G$
    \item[(3)] $d_{\text{GBC}}^{(R)} = G$
\end{itemize}
Here, (2) and (3) make sure that players do not loose their good reputation when someone cooperates with them. 
(1) can be interpreted as forgiveness without apologizing.
Here, players with a bad reputation can gain a good reputation if someone with a good reputation defects against them. We refer to these social norms as \textbf{forgiving}.
Slightly lower levels of cooperation can be achieved for $d_{\text{BGD}}^{(R)} = B$. 
In this case, apologies are still needed in order for bad players to be forgiven and this ensures that AllD players cannot invade the population as easily. We refer to these social norms as \textbf{unforgiving}.
Taylor expansion of the level of cooperation and benefit-to-cost ratios reveal this inherent trade-off between forgiving and unforgiving social norms. 
If players are more forgiving ($d_{\text{BGD}}^{(R)} = G$), they can ensure higher levels of cooperation. 
However, the benefit of cooperation needs to be higher for cooperation to be sustainable. 
On the other hand, if players are less fogiving, they prevent the invasion of cheaters, but levels of cooperation also decrease. 
For the results of the Taylor expansion analysis, see Table \ref{tb:taylor}. Additionally, for an extensive discussion of the ESSes when $q=\frac{1}{2}$ see \cite{Murase2023}.

\begin{table}[tb]
 \begin{tabular}{|c | c | c |} 
 \hline
 \textbf{Recipient Norm}& \textbf{Cooperation Level} & \textbf{Benefit-Cost-Ratio}\\ [0.5ex] 
 \hline\hline
 \{* G * B, * * G G\}& $1-\mu_a - 2\mu_4 + \mathcal{O}(\mu^2)$& $1/q + \mathcal{O}(\mu/q)$  \\
 \{* G * G, * * G G\}& $1-\mu_a - (q+1)\mu_4 + \mathcal{O}(\mu^2)$& $1/q + \mathcal{O}(\mu/q)$  \\
 \{* B * B, * * G G\}& $1-\mu_a/q - (q+1)/q \mu_4 + \mathcal{O}(\mu^2)$& $1 + \mathcal{O}(\mu/q)$  \\
 \{* B * G, * * G G\}& $1-\mu_a/q - 2\mu_4 + \mathcal{O}(\mu^2)$& $1 + \mathcal{O}(\mu/q)$  \\
 \hline
 \end{tabular}
 \caption{Taylor expansion for all highly cooperative ESS (CESS) social norm/behavioral strategy pairs.} \label{tb:taylor}
 \end{table}

\begin{figure}[tb]
    \centering
    \includegraphics[width=\textwidth]{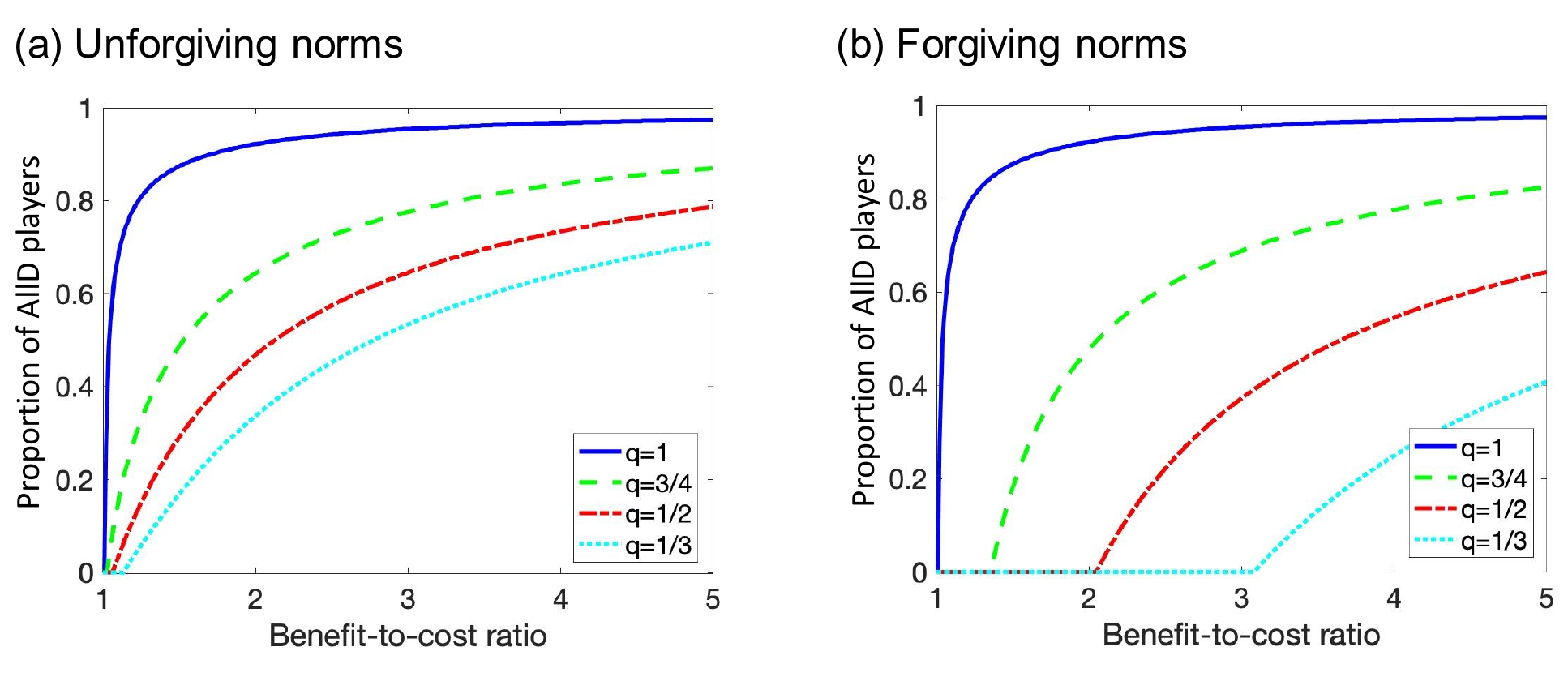}
    \caption{The basin of attraction for the two kinds of social norms. Forgiving social norms require higher b/c ratio, but achieve higher levels of cooperation. Model parameters are: $B=2, C=1, \mu_e = \mu_a = 0.01$}
    \label{fig:boa}
\end{figure}

An important factor when analyzing these ESSes is their ability to persist against AllD players. When comparing the basin of attraction for different values of $q$, in Table \ref{tb:taylor} we see that the benefit-cost-ratio necessary for cooperation to persist depends on $q$, in particular for forgiving social norms. For a further illustration of this, we compare the basin of attraction for forgiving and unforgiving social norms. An illustration of this can be seen in Figure \ref{fig:boa}. Here, we see how unforgiving norms have a larger basin of attraction that forgiving norms. However, both converge to 0 as $q \to 0$.

While this work is similar to \cite{Murase2023}, it introduces a new parameter: the time scale of updating $q \in (0,1)$. 
In particular, while \cite{Murase2023} discusses the cooperation levels (and additional social norms) in more detail, we generalize their framework and discuss in more detail the underlying mechanism behind the trade-off of forgiving. 
Our time-scale parameter $q$ builds the basis of the next section.

\subsection{A third time-scale governing recipient norms}

While the first part of this work focuses on the emergence of cooperation from defection, the next part focuses on comparing different social norms. In particular, we compare different time-scales $q$, i.e. rates to update recipient norms. To compare the different $q$-values, we introduce different gossip groups of players who share the same social norm and opinion about other players. We make similar assumptions as in the base model: (a) an indirect observation model and (b) an infinite population. Then, we assume that proportion $a$ of the population updates the donor's reputation with probability $q_1$, while proportion $(1-a)$ of the population updates the donor's reputation with rate $q_2$.

In this model, the reputation dynamics are given as follows:

\begin{align*}
    \dv{h_{G,G}^{(n)}(t)}{t} &= \sum_{i,j = G,B} h_{i,j}^{(n)}(t) \sum_{k,l = G,B} h_{k,l}(t) d_{i,k}(q_1) d_{j,l}(q_2) \\
    \dv{h_{G,B}^{(n)}(t)}{t} &= \sum_{i,j = G,B} h_{i,j}^{(n)}(t) \sum_{k,l = G,B} h_{k,l}(t) d_{i,k}(q_1) (1-d_{j,l}(q_2)) \\
    \dv{h_{B,G}^{(n)}(t)}{t} &= \sum_{i,j = G,B} h_{i,j}^{(n)}(t) \sum_{k,l = G,B} h_{k,l}(t) (1-d_{i,k}(q_1)) d_{j,l}(q_2) \\
    \dv{h_{B,B}^{(n)}(t)}{t} &= \sum_{i,j = G,B} h_{i,j}^{(n)}(t) \sum_{k,l = G,B} h_{k,l}(t) (1-d_{i,k}(q_1)) (1-d_{j,l}(q_2)) \\
\end{align*}
with $$h_{G,G}^{(1)}(t)+h_{G,B}^{(1)}(t)+h_{B,G}^{(1)}(t)+h_{B,B}^{(1)}(t)=1$$ and $$h_{G,G}^{(2)}(t)+h_{G,B}^{(2)}(t)+h_{B,G}^{(2)}(t)+h_{B,B}^{(2)}(t)=1$$
Here, we denote by $h^{(n)}_{i,j}(t)$ the proportion of players from group $n$ with reputation $i$ in the resident group and reputation $j$ in the mutant group at time $t$. Further, $h_{i,j}(t) = a h^{(1)}_{i,j}(t) + (1-a) h^{(2)}_{i,j}(t)$. We denote the equilibria of the reputation dynamics as $h^{(n)}_{i,j}$.

Then, the probability of a player of each group cooperating with a player of group $j$ is given as 
\begin{align*}
    \theta(1,j) &= p_{\text{GG}} (h^{(1)}_{\text{GG}}+h^{(1)}_{\text{GB}})(h^{(j)}_{\text{GG}}+h^{(j)}_{\text{GB}}) + p_{\text{GB}} (h^{(1)}_{\text{GG}}+h^{(1)}_{\text{GB}})(h^{(j)}_{\text{BG}}+h^{(j)}_{\text{BB}}) \\
    &+ p_{\text{BG}} (h^{(1)}_{\text{BG}}+h^{(1)}_{\text{BB}})(h^{(j)}_{\text{GG}}+h^{(j)}_{\text{GB}}) + p_{\text{BB}} (h^{(1)}_{\text{BG}}+h^{(1)}_{\text{BB}})(h^{(j)}_{\text{BG}}+h^{(j)}_{\text{BB}}) \\
    \theta(2,j) &= p_{\text{GG}} (h^{(2)}_{\text{GG}}+h^{(2)}_{\text{BG}})(h^{(j)}_{\text{GG}}+h^{(j)}_{\text{BG}}) + p_{\text{GB}} (h^{(2)}_{\text{GG}}+h^{(2)}_{\text{BG}})(h^{(j)}_{\text{GB}}+h^{(j)}_{\text{BB}}) \\
    &+ p_{\text{BG}} (h^{(2)}_{\text{GB}}+h^{(2)}_{\text{BB}})(h^{(j)}_{\text{GG}}+h^{(j)}_{\text{BG}}) + p_{\text{BB}} (h^{(1)}_{\text{GB}}+h^{(1)}_{\text{BB}})(h^{(j)}_{\text{GB}}+h^{(j)}_{\text{BB}}) \\
\end{align*}
Consequently, the payoff for players in each group are
\begin{align*}
    W_1(q_1,q_2|a) &= a \theta(1,1) (B-C) + (1-a) \theta(2,1) B - (1-a) \theta(1,2) C \\
    W_2(q_1,q_2|a) &= (1-a) \theta(2,2) (B-C) + a \theta(1,2) B - a \theta(2,1) C
\end{align*}

Numerical simulations suggest that close to $q_1 \approx q_2$, the payoff does not depend on $a$, as long as $1>a>0$. Specifically, if we define
\begin{align*}
    D(q|a) = \left. \pdv{(W_2(q_1,q_2|a)-W_1(q_1,q_2|a))}{q_2}\right|_{q_1=q_2=q},
\end{align*}
then $D(q|a_1)=D(q|a_2),\forall\, 0<a_1,a_2<1$. Note here that this does not extend to $a=0$, i.e. the appearance of a single invader who does not influence the dynamics of the remaining population as discussed for the dynamics in the previous subsection. 

As for the adaptive dynamics approach \cite{Geritz1997}, under some assumptions (large population size, small mutations), the evolution of $q$ is governed by $D(q)$. If $D(q) > 0$, then $q$ will increase. If $D(q) < 0$, $q$ will decrease. The population moves towards the points $q^*$ where $D(q^*)=0$ and $D'(q^*) <0$ and away from the points $q^*$ where $D(q^*)=0$ and $D'(q^*) >0$.

In this model, we can distinguish two different cases: 
\begin{enumerate}
    \item $d_{\text{GBC}}^{(D)} = B$
    \item $d_{\text{GBC}}^{(D)} = G$
\end{enumerate}
Here, whether we obtain a good reputation for cooperating with bad players significantly impacts the dynamics.

\begin{figure}[tb]
    \centering
    \includegraphics[width=\textwidth]{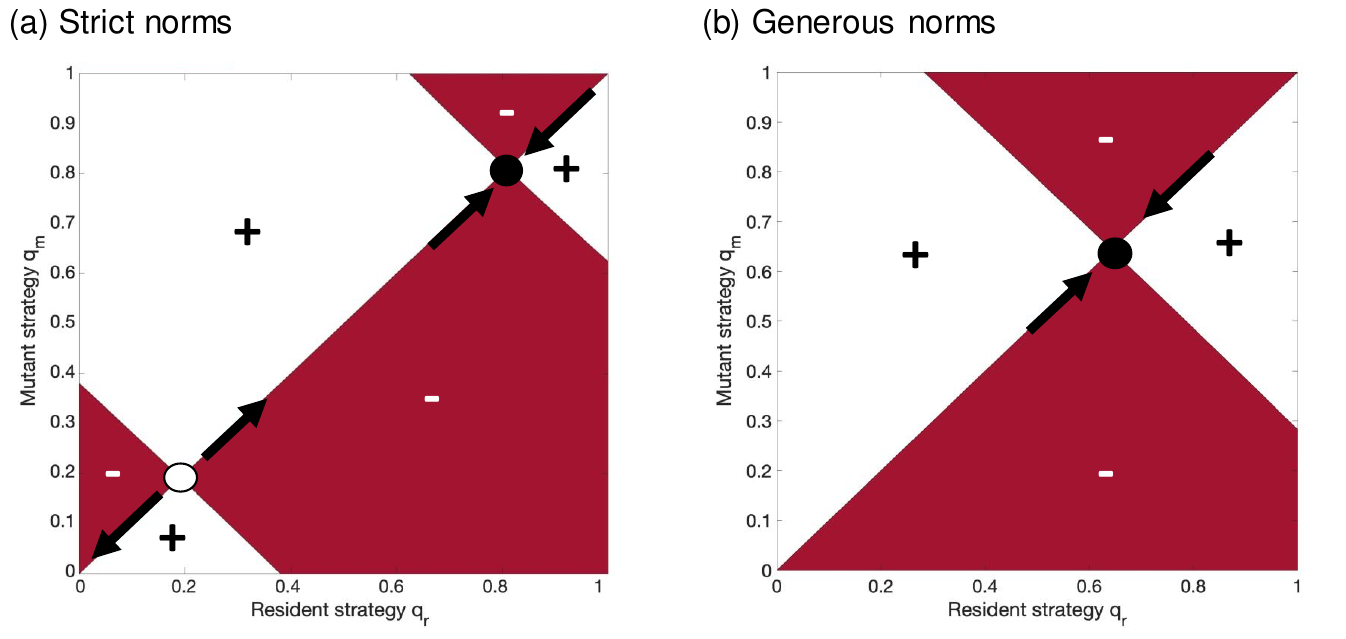}
    \caption{Pairwise invasibility plot for two different kinds of social norms.}
    \label{fig:qevo}
\end{figure}

In case (1) -- subsequently refered to as \textbf{strict social norms} -- players obtain a bad reputation if they cooperate with bad players. Here, two singular strategies exist: a smaller repellor $0< q_1^*$ and a larger attractor $q_1^*< q_2^* < 1$. If the population starts with $q > q_1^*$, it converges to $q_2^*$. Players care about a player's actions as well as other players' actions towards them to some extend. Once we pass the threshold $q < q_1^*$, however, $q$ converges to $q=0$. Here, the benefit of cooperation necessary to maintain cooperation converges to $b/c \to \infty$ and cooperation is not sustainable anymore. Cheaters (AllD) will invade cooperative strategies. An illustration of these results can be seen in Figure \ref{fig:qevo} (a).

In case (2) -- subsequently refered to as \textbf{generous social norms} -- players keep their good reputation if they cooperate with bad players. In this case, there exist a single global attractor and the population converges to the attractor $q_2^* \in (0,1)$. For an illustration, see Figure \ref{fig:qevo} (b).

We focus on OR players applying strategy $p=\begin{pmatrix}
    0 & 1 \\ 0 & 1
\end{pmatrix}$. Note that 
$$
    \pdv{q_2}\left(h_{\text{BG}}^{(2)} + h_{\text{GB}}^{(1)} \right) = \pdv{q_2}\left(h_{\text{BG}}^{(1)} + h_{\text{GB}}^{(2)} \right).
$$
We define $G^{(i)}_{j}$ as the proportion of good players in group $i$ as perceived by group $j$. Therefore, 
\begin{align*}
    0 &= \pdv{}{a}\left( (1-a)(B-C) \pdv{G_2^{(2)}}{q_2}  - a (B-C) \pdv{G^{(1)}_1 }{q_2} + aB \pdv{G^{(1)}_2 }{q_2} \right. \\ & \left. +(1-a)C \pdv{G^{(1)}_2 }{q_2} - (1-a)B \pdv{G^{(2)}_1 }{q_2} - a C \pdv{G^{(2)}_1 }{q_2}\right) \\
    &= \pdv{}{a}\left( \left.\pdv{(W_2(q_1,q_2|a)-W_1(q_1,q_2|a))}{q_2} \right|_{q_1=q_2=q}\right) \\
    &= \pdv{D(q|a)}{a}
\end{align*}
and 
\begin{align*}
    D(q) & =(B-C)\pdv{G^{(2)}_2 }{q_2}-B\pdv{G^{(2)}_1 }{q_2}+C\pdv{G^{(1)}_2 }{q_2} \\
    &= B \pdv{q_2} \left. ( h_{\text{GG}}^{(2)} + h_{\text{BG}}^{(2)} - h_{\text{GG}}^{(1)} - h_{\text{BG}}^{(1)} ) \right|_{q_1=q_2=q}- \left. C \pdv{q_2} ( h_{\text{BG}}^{(2)} - h_{\text{GB}}^{(2)})\right|_{q_1=q_2=q}
\end{align*}
does not depend on $a$. We can apply a similar argument to CO-players using $p=\begin{pmatrix}
    1 & 1 \\ 0 & 1
\end{pmatrix}$.

\section{Discussion}

Our exploration into the nuances of indirect reciprocity builds upon a large body of prior work. Using the leading eight model \cite{Ohtsuki2004}, we have further built upon the work on dual reputation updates \cite{Murase2023} by introducing a third time-scale parameter that regulates how frequently reputations are updated based on individuals' own actions compared to other player's actions. With dual reputation updates, we have introduced a trade-off: Forgiving social norms do not expect an apology of 'bad' players. Once someone has been punished, they obtain a good reputation. These norms can obtain high levels of cooperation, but need a large benefit-to-cost ratio. Unforgiving social norms do expect an apology from a 'bad' player. They obtain slightly lower level of cooperation, but can be sustained for smaller benefit-to-cost ratios. Additionally, they have larger basins of attraction, especially for large emphasize on the recipient updates.

We have further built upon previous work comparing the leading eight \cite{Kessinger2023} to compare the different time-scale parameters. This establishes two cases of different social norms: Strict social norms that punish 'good' players that cooperate with 'bad' players and generous social norms that do not punish 'good' players for cooperating with 'bad' players. For strict social norms, the population might end up in an equilibrium where $q=0$. In this case, defectors can invade the population and cooperation collapses. Generous social norms, on the other hand, have a single attracting equilibrium with $q > 0$. Here, cooperation can be sustained for high enough benefit-to-cost ratio.

Our findings contribute to the discourse on indirect reciprocity by illustrating the critical importance of how communities assess interactions between cooperative individuals and those perceived negatively. The model derived in \cite{Ohtsuki2004} suggests that the updates of reputations in scenarios where good individuals cooperate with bad ones have negligible consequences on the level and evolutionary stability of cooperation. Our model shows that evaluating good players cooperating with bad players can have significant consequences for the persistence of cooperation. Strict norms that punish good players for this form of cooperation can lead to the collapse of cooperation through cheaters while generous norms that do not punish them can preserve cooperation.

We note here that both, indirect observation and this separation into three time-scales are significant assumptions that are not necessarily true. 
In particular, the indirect observation assumption has been challenged and the introduction of private information has been found to have profound impacts on the dynamics as discussed in (see \cite{Hilbe2018,Schmid2023}), though relaxing the assumption of binary reputations can stabilize the behavior. The significant interest in direct observation models and private reputations \cite{Hilbe2018,Uchida2010,Ohtsuki2015,Schmid2023}, shows the importance of moving beyond the indirect observation model used in this work. Additionally, we have relied in binary reputations and it would be interesting to include a more nuanced reputation model as has been done in \cite{Murase2022,Lee2021}. For the extension of the model, we have made additional assumptions. In particular, we have assumed that the populations are well-mixed for the prisoner's dilemma interactions, but for 'gossip exchange'. Future work should adress this and introduce some form of intermediate mixing for each scenario. 

\backmatter

\bigskip









\bibliography{sn-bibliography}

\end{document}